\documentclass[isre,nonblindrev]{informs4}
\usepackage{hyperref}
\hypersetup{
    colorlinks = true,
    urlcolor = {blue},
    citecolor= [rgb]{0,0.5,0.5}
}
\usepackage{cleveref}
\usepackage{graphicx}
\usepackage[flushleft]{threeparttable}
\usepackage{footnote}
\usepackage{multirow}
\usepackage{booktabs}
\usepackage{soul,xcolor}
\usepackage{titlesec}
\usepackage{appendix}
\usepackage{verbatim}
\usepackage{arydshln}
\usepackage{ebgaramond}
\usepackage{pdflscape}
\usepackage{colortbl,hhline}
\usepackage{longtable}
\usepackage{arydshln}
\usepackage{subcaption}
\usepackage[sort,comma]{natbib}
\usepackage{amsmath, bm}

\DeclareMathAlphabet{\mathsfit}{T1}{\sfdefault}{\mddefault}{\sldefault}
\SetMathAlphabet{\mathsfit}{bold}{T1}{\sfdefault}{\bfdefault}{\sldefault}

\usepackage{bbm}

\DoubleSpacedXI

\usepackage{natbib}
 \bibpunct[, ]{(}{)}{,}{a}{}{,}%
 \def\bibfont{\small}%

\TheoremsNumberedThrough     % Preferred (Theorem 1, Lemma 1, Theorem 2)
\ECRepeatTheorems
\EquationsNumberedThrough    % Default: (1), (2), ...
\MANUSCRIPTNO{MS-0000-0000.00}

\titleformat{\subsubsection}
  {\normalfont\normalsize\bfseries}{\thesubsubsection}{1em}{}
\titlespacing*{\subsubsection}{0pt}{3.25ex plus 1ex minus .2ex}{0ex plus .2ex}

\begin{document}

\RUNTITLE{}

\TITLE{Unraveling Human–AI Teaming: A Review and Outlook\thanks{The authors are named in alphabetical order. All authors contributed equally.}}
%\TITLE{Unraveling Human–AI Teaming: A Review and Outlook}

\ARTICLEAUTHORS{%
\AUTHOR{Bowen Lou}
\AFF{University of Southern California, \EMAIL{bowenlou@marshall.usc.edu}}
\AUTHOR{Tian Lu}
\AFF{Arizona State University, \EMAIL{lutian@asu.edu}}
\AUTHOR{T. S. Raghu}
\AFF{Arizona State University, \EMAIL{Raghu.Santanam@asu.edu}}
\AUTHOR{Yingjie Zhang}
\AFF{Peking University, \EMAIL{yingjiezhang@gsm.pku.edu.cn}}

}

\ABSTRACT{%
Artificial Intelligence (AI) is advancing at an unprecedented pace, with clear potential to enhance decision-making and productivity. Yet, the collaborative decision-making process between humans and AI remains underdeveloped, often falling short of its transformative possibilities. This paper explores the evolution of AI agents from passive tools to active collaborators in human-AI teams, emphasizing their ability to learn, adapt, and operate autonomously in complex environments. This paradigm shifts challenges traditional team dynamics, requiring new interaction protocols, delegation strategies, and responsibility distribution frameworks. Drawing on Team Situation Awareness (SA) theory, we identify two critical gaps in current human-AI teaming research: the difficulty of aligning AI agents with human values and objectives, and the underutilization of AI’s capabilities as genuine team members. Addressing these gaps, we propose a structured research outlook centered on four key aspects of human-AI teaming: formulation, coordination, maintenance, and training. Our framework highlights the importance of shared mental models, trust-building, conflict resolution, and skill adaptation for effective teaming. Furthermore, we discuss the unique challenges posed by varying team compositions, goals, and complexities. This paper provides a foundational agenda for future research and practical design of sustainable, high-performing human-AI teams.
}%

\KEYWORDS{Artificial Intelligence; AI Agent; Human–AI Teaming; Situation Awareness; Research Outlook}

\maketitle
\thispagestyle{empty}

\DoubleSpacedXI

\newpage
\setcounter{page}{1}
\section{Introduction}
\label{sec:intro}

Artificial Intelligence (AI) is advancing at an unprecedented pace, transforming industries and reshaping daily life. From healthcare and finance to education and manufacturing, AI-based systems are increasingly supporting humans across various fields \citep{lu20221+, lou2021ai, berente2021managing}. Yet, despite AI’s potential, collaborations between humans and AI remain ad hoc, often resulting in underwhelming performance \citep{teaming2022state}. The promise of AI lies in its ability to integrate into teams to augment human capabilities, improve decision-making, enhance productivity, and drive innovation. Realizing the benefits, however, requires a more systematic examination of the novelties associated with human-AI teaming.

In traditional human-human teaming, researchers have extensively studied shared awareness, communication, and collaboration, establishing frameworks around communication frequency, shared mental models, task complexity, and training. Similarly, human-system teaming has focused on the use of communication and collaboration tools, which support team coordination but remain distinct from the tasks themselves \citep{desanctis1987foundation,jerry2000group,dennis2002investigating}. Implicit in all team-based interactions is the assumption that team members share common values and work together toward shared goals and objectives. The perspective that technology is a tool to enhance human efforts needs a refresh with the emergence of AI agents. Recent studies have shown that AI agents are evolving rapidly, with their capabilities doubling approximately every seven months since 2019 \citep{kwa2025measuring}.\footnote{The concept of AI agents is not new. \cite{russell2016artificial} define them as entities that perceive their environment and act rationally. This paper builds on and modernizes that foundational view to reflect the advanced capabilities of today’s AI. Today's AI agents have evolved into autonomous, iterative, and intelligent team members. They learn, reflect, and adapt within complex and dynamic environments. Moreover, they are no longer passive tools but active collaborators in human-AI teams in sustained interaction, emotional intelligence, and autonomous problem-solving. This marks a paradigm shift beyond the classical notion of rational action.} In this regard, human-AI teaming introduces a new era of team science, requiring revised interaction protocols for AI agents, and new approaches to delegation, task execution, and responsibility sharing.

In this paper, we examine the emerging landscape of human-AI teaming, where AI has moved beyond simple automation to become an intelligent, agentic artifact \citep{dennis2023ai, gao2024empowering}. The rapid rise of large language models (LLMs) such as Generative Pre-trained Transformer (GPT) has accelerated this transformation. AI-enabled agentic artifacts are no longer simply embedded into workflows; they are now intelligent agents capable of actively taking on mission-critical roles and tasks within teams  \citep{mckinsey2023economic, mckinsey2024why, alavi2024genairedefine}. These systems learn from interactions, adapt to changing contexts, and make independent decisions, transforming AI into an autonomous partner in collaborative efforts. The agentic shift highlights some gaps in the current information systems (IS) literature, which mainly treats AI as a tool rather than a nearly full-fledged teammate \citep{wang2019friend, dennis2023ai, lu20221+}.

We identify two major research gaps. First, there is a significant challenge in designing AI agents that align with human values and objectives. This gap arises because humans themselves often struggle to explicitly articulate their goals and objectives.  Effective team protocols and processes must account for these ambiguities to enable AI agents to augment human capabilities. Advancing human-AI collaboration requires strategies that bridge this gap and facilitate meaningful integration. 
The second gap stems from the limited integration of AI’s full capabilities into collaborative workflows. Many still view AI agents as static tools rather than adaptive, knowledge-enhancing partners. This narrow perspective restricts AI’s potential in dynamic decision-making and strategic reasoning while behavioral resistance and inertia further reinforce the issue \citep{polites2012shackled}. Many individuals hesitate to accept AI as a genuine team member, leading to underutilization and even adverse outcomes. Bridging this gap requires reconceptualizing AI as an active partner in knowledge-intensive tasks, fostering deeper interaction between human and algorithmic intelligence to unlock AI’s full collaborative potential.

This paper aims to bridge these gaps by applying the well-established Team Situation Awareness (SA) theory to the context of human-AI teaming. We argue that integrating AI agents into teams not only requires designing systems that can work toward shared objectives but also requires a rethinking of how teams execute tasks alongside AI agents. By doing so, we develop an extended model of Team SA in human-AI teaming, and propose a structured outlook for future research. In particular, we propose a systematic framework for understanding human-AI teaming across four key dimensions: team formulation, coordination, social dynamics, and knowledge creation. First, effective team formulation requires shared mental models that align AI’s decision-making with human values and objectives, ensuring clarity in role specification while allowing flexibility for adaptation. Team coordination focuses on optimizing task delegation and human-AI interaction protocols, minimizing expectation misalignments and ensuring interpretable and reliable decisions. Team maintenance is crucial for trust-building, conflict resolution, and sustaining high-performance collaboration, particularly in managing AI’s limitations and accountability in decision-making. Finally, team training and evolution involve continuous learning mechanisms that enable AI to refine its capabilities while ensuring human members continue to create new knowledge. By structuring research around these four interconnected areas, we outline a roadmap for advancing human-AI teaming and addressing critical challenges such as trust, adaptability, accountability, and long-term sustainability in AI-integrated teams.

\section{Revisiting Human-AI Teaming}
\label{sec:teaming}

\subsection{Emerging Agentic AI Systems}
\label{sec:intelligentagentdef}

Information systems are beginning their transformation into agentic artifacts that ``are no longer passive tools waiting to be used, are no longer always subordinate to the human agent, and can now assume responsibility for tasks with ambiguous requirements and for seeking optimal outcomes under uncertainty'' \citep{baird2021next}. Modern AI systems, with their ability to ``learn and act'' independently of human instructions, are showing the promise to emulate human thought processes, behavior, and decision-making \citep{dennis2023ai, durante2024agent}. Although these systems may remain within the realm of Narrow AI, their capabilities now extend to many tasks that, until recently, required highly trained professionals.

Generative AI systems, such as ChatGPT, Midjourney, Claude, Llama, and Gemini, excel in tasks requiring nuanced judgment and adaptability --- areas where classical AI systems often falter due to ``brittleness" when encountering scenarios outside their predefined rules \citep{mckinsey2023economic, dhar2023paradigm}. They now assist in domains previously dominated by human cognition and expertise \citep{zhou2024generative, susarla2023janus}. For example, ChatGPT interprets ambiguous prompts to draft complex documents and hold meaningful conversations. Similarly, Midjourney generates original artwork from simple text, demonstrating creative problem-solving skills that go beyond rule-based automation. Importantly, these capabilities pave the way for the emergence of agentic AI systems—systems that not only assist but also act autonomously and collaborate dynamically with humans \citep{durante2024agent}. Agentic AI engages in creative, coordinated tasks, takes initiative in decision-making, and adapts through iterative feedback. By generating context-aware outputs and exhibiting independent reasoning, they challenge traditional models of human-AI interaction.

The key characteristics of these agents include:
\begin{itemize}
    \item \textbf{Planning}: The ability to plan and execute multi-step processes to achieve requested goals. For example, an agent can outline an essay, conduct online research, and draft the essay in a sequential manner \citep{ng2023how}.
    \item \textbf{Self-explanation}: A notable feature that allows the agent to communicate and reason effectively, possibly requiring a robust theory of mind for both itself and the human users \citep{bubeck2023sparks}.
    \item \textbf{Sustained interaction}: The facilitation of dynamic, prolonged engagements with users, moving beyond static, single-question interactions \citep{traumer2017towards}.
    \item \textbf{Reflection}: The capacity to reflect on their work to identify and implement improvements, thereby continually enhancing their performance \citep{ng2023how}.
    \item \textbf{Prescription/autonomous functioning}: The ability to function as autonomous work partners, by prescribing or taking actions, and/or collaborating with human users to achieve common objectives \citep{baird2021next,guzman2020artificial}. 
\end{itemize}

These characteristics enable agents to undertake sequential tasks with multiple iterations alongside human users, demonstrating chain-of-thought reasoning and providing explanations for their outputs. Continuous improvements in these abilities could eventually lead to the deployment of sophisticated agents that autonomously handle tasks alongside human users. Agentic AI systems are fulfilling domain-specific tasks such as web search, code execution, data processing, customer support, and other professional services \citep{rahwan2019machine}. Their ability to learn from diverse data sources and generalize to new and unforeseen scenarios positions them for even broader applications in organizational and team-based processes. In this context, these intelligent agents have been conceptualized as implicit computational models of humans—referred to as ``homo silicus'' \citep{filippas2024large}. Notably, their capability is not merely limited to resolving problems in a virtual environment (i.e., cyberspace) but also navigating the complexity and unpredictability of the physical world \citep{xu2024survey, liu2024aligning}. In the following sections, we interchangeably use the terms ``intelligent agents'', ``agents'', and ``AI'' (in contrast to ``traditional AI'') to refer to such systems that possess the aforementioned intelligent characteristics.

\subsection{Definition of Human–AI Teaming}
\label{sec:teamingdef}

According to \citet{salas1992toward}, a team is a group of two or more individuals who share common goals, role assignments, and interdependence. Typically, team dynamics (i.e., teaming) involve task completion within a specific context, application of specialized knowledge and skills, and performance under time, space, resource, cognitive, and other constraints. Teamwork includes two categories of behavior: \textit{task-work} and \textit{team-work}. Task-work involves actions performed by individual team members essential for carrying out their specific functions. Teamwork involves knowledge, skills, and attitudes that \textit{facilitate} interaction among team members, ensuring coordination to achieve team goals \citep{morgan1986measurement}. Essentially, teams operate as socio-technical systems, where well-defined roles (akin to societal roles) and effective coordination mechanisms are fundamental to achieving high performance, especially in high-stakes environments and complex tasks \citep{salas1992toward}.

To initiate a team, it is necessary to confirm participant configurations, superordinate goals and priorities, the interdependence of teammate goals, sources of information and instruction for tasks, as well as factors such as team cohesion, communication, and coordination \citep{salas1992toward}. At its core, teamwork is a form of collaboration. Regardless of the levels of collaboration \citep{aaltonen2018refining}, an essential characteristic of any team, including human-AI teams, is the presence of shared goals and the ability to complement each other in working on tasks (i.e., task assemblage) to achieve those goals \citep{rai2019next}.

\textit{Role specification} and \textit{role fluidity} are crucial complementary foundations for effectively running a team, balancing structure and adaptability to ensure seamless team formulation, coordination, and maintenance \citep{rousseau2006teamwork}. Role specification refers to the assignment of specific, well-defined roles to individuals based on expertise or task requirements. It provides structure and clarity, supporting effective team formulation and coordination \citep{tyler1973measuring}. Role fluidity, on the other hand, emphasizes flexibility and the interchangeability of roles within a team. It fosters adaptability and collaboration, enhancing team dynamics and guiding knowledge creation in dynamic environments and complex tasks \citep{faraj2006coordination, edmondson2001disrupted}. 
\Cref{sec:teamSA} elaborates more theoretical discussions on this point.

Agentic AI systems create new collaboration dynamics in teams. The infusion of such agents requires more cognitive input and coordination among human team members to ensure role fluidity and teamwork effectiveness \citep{gorman2014team}. To facilitate teamwork between humans and intelligent agents, all participants need to adapt to changes in role specification with the partnered agents \citep{zhang2021ideal}. Typically, human-AI teaming involves humans and intelligent agents interacting as team members in a context where the agent performs similar task-work and team-work functions as a human team member \citep{mcneese2021team}. Based on the previous discussions, we present a formal definition of human-AI teaming when the focal intelligent agents possess most of the characteristics defined in Section \ref{sec:intelligentagentdef}. \textit{In human-AI teaming, humans and intelligent agents collaborate to achieve shared goals. Human and AI team members dynamically coordinate their actions to plan, facilitate, iterate, and evaluate.\footnote{This follows \citet{gao2024taxonomy} who define the interaction flow of human-LLM co-working.}}

Human-AI teaming shares many similarities with human-human teaming, allowing for the extrapolation of existing interpersonal team management literature to this AI-related context. For example, both forms of teaming involve shared objectives. Each member, whether human or AI, can be assigned specific tasks, contributing to both task-work and team-work. Teamwork involves multiple iterative interactions and feedback loops \citep{peisaving2024saving}. Moreover, mutual learning is a key component and driver of team performance.

There are, however, several key factors that distinguish AI agents from typical human teammates or traditional technical tools. While AI agents can be subject to human control in terms of task assignments and (sub)objectives, their advanced capabilities can disrupt traditional team hierarchies by flattening authority structures and redistributing expertise \citep{tambe2025ai, wu2019data}. In addition, unlike human teammates, AI agents impose less social and emotional burden, as they do not impose the same level of interpersonal skills requirement as in human teamwork. This lack of emotional involvement can alter the dynamics within a team, reducing emotional support and potentially affecting cohesion \citep{xu2023heterogeneous}. AI agents may remain more compliant in conflicts, lacking the nuanced emotional responses that humans might offer in tense situations. A further distinction lies in the question of accountability. Unlike human teammates, AI cannot be entirely responsible for decisions or actions, which raises questions about how responsibility is distributed within a team. This ambiguity can complicate trust-building and decision-making processes within a team. Moreover, AI's potential for deceptive (possibly due to issues such as hallucination \citep{banerjee2024llms}) or overly accommodating behavior adds another layer of complexity, as it can sometimes cater to user expectations in misleading ways, creating false confidence and undermining the reliability of collaboration. These challenges call for a new theoretical understanding of human-AI teaming factors that contribute to achieving shared objectives.

\subsection{Theoretical Framework: An Extended Model of Team Situation Awareness}
\label{sec:teamSA}

\begin{figure}[t]
\centering
    \includegraphics[trim={32mm 50mm 40mm 37mm}, clip, width=\textwidth]{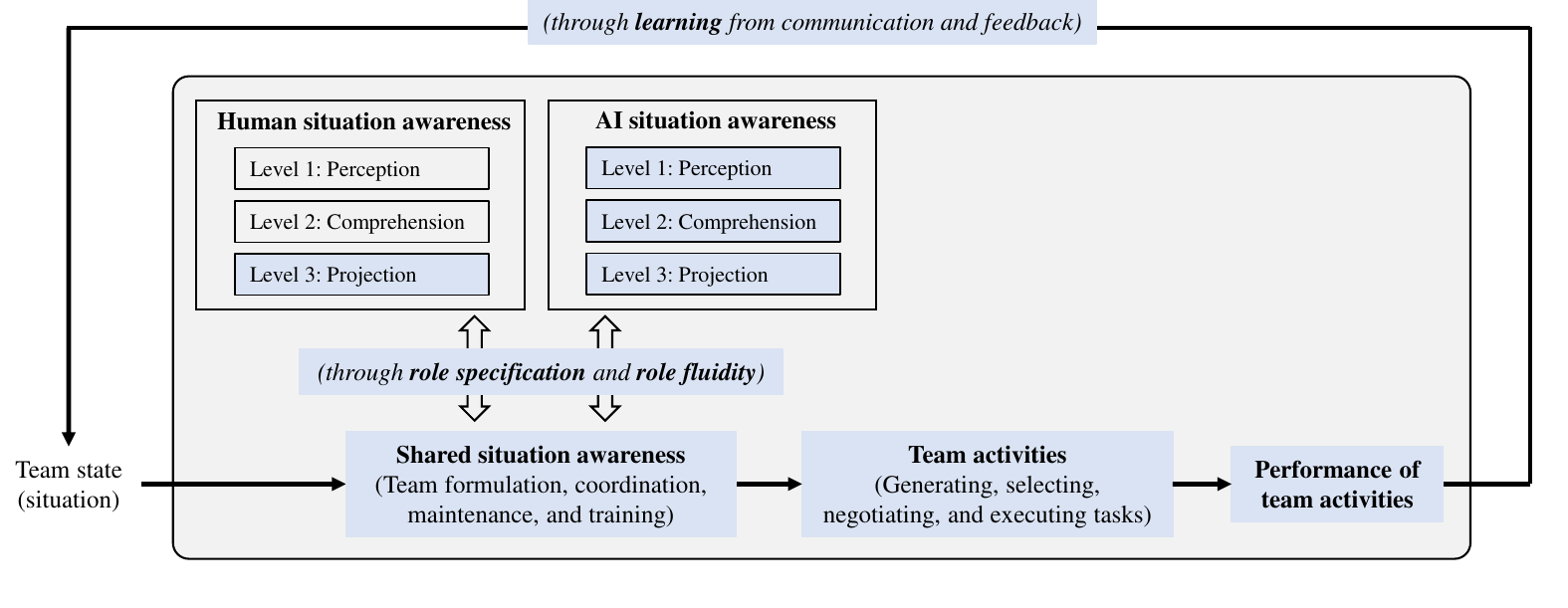}
\begin{tablenotes}
      \footnotesize
      \item \textit{Note:} The model is developed based on the literature of human SA \citep{endsley1995toward}, team SA \citep{salas2017situation, teaming2022state}, group-related activities \citep{mcgrath1982group} and action team learning \citep{vashdi2013can}. Blue-shadowed boxes indicate their pertinence to the major future directions to explore (\Cref{sec:outlooks}).
     \end{tablenotes}
\caption{An Extended Model of Team SA in Human–AI Teaming}
\label{fig:teamSA}
\end{figure}

The classic team situation awareness (team SA) theory \citep{endsley1995toward} provides a foundational framework for understanding the intricate dynamics (i.e., teaming process) that shape effective teamwork. As illustrated in \Cref{fig:teamSA}, a basic element of the team SA model is individual SA, which includes three levels: \textit{perception}, \textit{comprehension}, and \textit{projection}. Individual perception and comprehension SA enable team members to interpret and process information related to the collaborative environment. Comprehension SA enables understanding of teammate's task contributions. Projection involves anticipating future actions of team members and consequences \citep{endsley1995toward}. 

In addition to individual SA, the team SA model emphasizes the critical concept of ``shared SA'' \citep{teaming2022state}, defined as ``the degree to which team members possess the same SA on shared SA requirements'' \citep{endsley2001model}. Shared SA is based on shared objectives among team members and goes beyond individual SA by encompassing the collective understanding of team members regarding the shared SA requirements. This requires cognitive investment and thus constitutes a ``shared mental model'' \citep{blickensderfer1997theoretical}. Such a model is crucial for successful collaboration. Specifically, in a teaming process, shared SA is crucial to foster effective and creative idea/planning generation, and streamline the team work process for task implementation \citep{straus1999testing}. Shared SA underscores the cohesive fabric that binds team members into a synchronized unit and minimize the potential for miscommunication and misinterpretation \citep{bubeck2023sparks}. 

Recognizing the participation of intelligent agents in a team \Cref{sec:teamingdef}, we extend the team SA model to incorporate role specification and role fluidity. Role specification ensures that all team members, including AI, clearly understand their specific tasks while effectively accommodating themselves with other's expertise. Moreover, teamwork thrives on adaptability, open communication, and shared responsibilities -- core functions of role fluidity. Flexible roles allow teams to respond effectively to dynamic and uncertain environments, fostering collaboration and innovation. 
Team learning is therefore crucial for achieving sustainable team performance \citep{vashdi2013can}. It also aligns with and underscores the inherent dynamism of shared SA in human-AI teaming. This dynamism is driven by continuous feedback loops, ensuring that shared SA remains adaptive and responsive to changing circumstances \citep{demir2023modeling}.

Team-related activities, as outlined in the Circumplex model of group tasks \citep{posner2005circumplex}, typically involve \textit{generating} ideas or plans, \textit{selecting} the most suitable solution, \textit{negotiating} a resolution to conflicts, and \textit{executing} tasks. In the context of human-AI teaming, it becomes crucial to revisit and explore how team SA interacts with these complex tasks. AI's capacity to rapidly process large volumes of data and generate insights can work for both role specification and fluidity to enhance team SA, potentially reshaping how teams handle decision-making, conflict resolution, and task execution. In addition, as discussed earlier, the integration of AI introduces new complexities, such as the need for alignment between human and AI-generated awareness. As AI's role in teams grows, revisiting the relationship between team SA and task performance is vital to ensure effective collaboration and to address potential gaps in mutual understanding.

\SingleSpacedXI
\DoubleSpacedXI

\section{Mapping Existing Works to Human-AI Teaming}
\label{sec:mapping}

We next briefly review relevant studies and integrate them into the extended team SA framework (\Cref{fig:teamSA}). 

\subsection{Human Perception SA of Intelligent Agent}

When humans and AI work together, Human \textit{perception SA} incorporates the capabilities of intelligent agents, recognizing their strengths, limitations, and relevant environmental cues. 

Early studies on the topic of human-AI teaming have documented a few benefits. AI or machines, in general, possess computational powers and might yield information and alternatives that humans do not consider on their own \citep{meyer2021impact}. In particular, when AI serves the role of assistants, as humans are flexible and adaptable to changing contexts, humans can learn from the machines that can possibly play an instructional role in improving the quality or accuracy of humans' decision-making \citep{choi2022does}, and overall welfare gains \citep{kleinberg2018human}. 
Humans might also welcome ceding control of certain decisions to AI. For example, humans may view algorithm-based task assignments as fairer than human-initiated task assignments \citep{bai2022impacts}. The ability of AI to process massive amounts of data for optimization is a contributor to the enhanced perception of SA.

Effective integration of human intelligence into agentic processes is still in its infancy. Humans possess distinct strengths that can contribute to better AI perception SA.  
The first strength is humans' domain expertise. In decision-making, individuals, especially those with years of experience, can draw on their institutional knowledge from multiple, often undocumented, sources -- commonly referred to as tacit knowledge \citep{cao2021man,choudhury2020machine,kumar2022data}. Intrinsic expertise allows humans to potentially outperform AI on perception SA.
Second, humans possess sophisticated emotional intelligence, knowledge of institutional politics, and industry insights, which are still difficult for AI to master \citep{ibrahim2021eliciting}. While humans can flexibly draw on these `private' sources of information to inform their decisions, this type of data is not available or feasible to use in training AI models.
Third, another key strength of humans is their interpersonal communication skills \citep{brynjolfsson2017can,luo2021artificial}. 

AI presents comparative advantages in routine-based, codifiable, or scripted tasks \citep{autor2015there}. There are, however, contexts where AI can contribute to better emotional responses as well. High-volume tasks create a cognitive overload on humans that places inherent limits on perception SA. Human ability to leverage intrinsic knowledge and emotional intelligence can be severely constrained during stressful situations, especially when multiple information sources have to be synthesized rapidly for decision-making. In such situations, AI can be more effective in formulating responses and augmenting human interpersonal communication skills \citep{luo2021artificial}. 
Effective human-AI teaming can complement each other's strengths to create an optimized perception SA. For example, human-AI teaming can leverage tacit knowledge, intuition, and empathy with rapid synthesis of codifiable knowledge and unstructured information \citep{feuerriegel2022bringing, lou2021ai}.
        
\subsection{Human Comprehension SA of Intelligent Agent}

\textit{Comprehension SA} emphasizes the importance of understanding how team decisions are derived. In human-AI teaming, this understanding prevents humans from passively accepting or rejecting AI contributions. Consequently, beyond the existing strengths of AI and humans, collaboration itself can create supermodular decision-making capability \citep{kawaguchi2021will}. Recent studies document and propose solutions to augment humans' comprehension SA. 
First, AI deployment evokes behavioral changes. On the one hand, AI frees humans from repetitive but boring tasks, which in turn allows them to allocate more cognitive resources to higher-level problem-solving \citep{jia2023and}. On the other hand, AI-enabled predictions or recommendations are additional sources of information for humans to comprehend, often without plausible explanations for recommendations \citep{liu2023unintended}. As such, human-AI teaming requires explicit designs to enhance comprehension SA. With intentional designs (e.g., showing risk scores \citealp{fogliato2022case}), humans can demonstrate better comprehension SA and feel empowered to overrule AI recommendations. 

Second, we could ideally expect humans to correct AI's mistakes in a timely manner, especially in ambiguous decision contexts. Prior studies show that when uncertain about their own decisions, especially when these conflict with AI's recommendations, humans actively gather more information, search for alternative cues, or employ heuristic reasoning \citep{jussupow2021augmenting}. However, this rethinking process is not guaranteed and depends on certain contextual factors, such as the presence of interesting and complex problems. Strategic rethinking requires a significant investment in cognitive resources, particularly when humans feel jaded performing routine work \citep{lebovitz2022engage}. 
Given the interchangeable roles of AI and humans in completing similar tasks, humans may find it easier to apply cognitive abilities beyond their typical domain expertise. Effective utilization of human-AI collaboration can thus generate new, sustainable sources of competitive advantage \citep{krakowski2022artificial}.

\subsection{Tasks in Human-AI Collaboration}

As discussed in Section \ref{sec:teamSA}, group tasks typically encompass four key types, yet existing literature predominantly focuses on the decision-making phase, where humans and AI collaborate to solve specific problems. This narrow focus overlooks other types of group tasks, such as negotiation, communication, and coordination tasks. In traditional human-AI collaboration, AI often takes on static, predefined roles as an assistant, primarily executing human assigned tasks. This dynamic reflects an imbalance, as humans retain decision-making authority while AI lacks agency, leading to an unequal partnership. On the one hand, when humans and AI choose solutions for the same tasks, the literature suggests and confirms the importance of keeping humans in the decision-making loop to enhance effectiveness. Human workers are more likely to accept AI assistance or even make contributions if they can manipulate or make adjustments to AI recommendations \citep{dietvorst2018overcoming,kawaguchi2021will}. In addition, researchers have examined how to incorporate human opinions (e.g., with voting strategy \citep{kamar2012combining}, through a reinforcement learning framework \citep{koster2022human}), when and how much to include human interventions \citep{de2020regression,lage2018human}, and even account for potential side effects of involving humans (e.g., generalizing rejection learning to mitigate biases held by external decision makers \citep{madras2018predict}). 
     
Many common processes in business require human-AI chaining in task design. That is, humans and AI may execute tasks sequentially, which requires both parties to understand and account for each other's capabilities. Therefore, designing appropriate task assignments is critical to optimizing joint performance. Prior studies suggest that collaborative augmentation is most effective when AI delegates tasks to humans, but not vice versa \citep{fugener2022cognitive,feuerriegel2022bringing}. Effective task allocation hinges on a comprehensive understanding of the complementary strengths of humans and AI, as well as well-designed delegation rules. However, humans may struggle to accurately assess their own capabilities, which can limit the effectiveness of the collaboration.
Notice that simply focusing on AI utilization might not work, as there exist systematic differences between perceived and actual benefits, whereas the latter contributes to the realization of collaborative values \citep{kim2022home}. Thus, firms should account for the potential difficulties human employees might face in executing AI's recommendations rather than merely encourage them to adopt AI as it is. 

The emergence of generative AI systems challenges traditional work allocations in human-AI collaboration, often emphasizing the complementary roles of humans and AI. Prior studies view humans as better suited for creative, intuitive, or strategic roles \citep{jia2023and}. Generative AI, however, blurs these distinctions by demonstrating substantial creative capabilities \citep{toner2024artificial,chen2024large}. Human-AI chaining of tasks will, therefore, need to be reassessed to account for generative AI's ability to contribute ideas, adapt to context, and collaborate more equitably. 

%\iffalse
\subsection{Human-human Teaming}
\label{sec:human-human}

Human characteristics play critical roles in shared SA and the ability to project teammates' actions in human-human teaming. Extant research highlights that individual traits—such as trust, empathy, and communication skills—are crucial for cultivating and sustaining team SA. For example, \cite{salas2017situation} demonstrate that trust among team members enhances information sharing and reduces hesitation in communication, leading to a more cohesive and synchronized awareness across the team. 
In addition, \citet{woolley2010evidence} find that collective intelligence—linked to social sensitivity, balanced communication, and gender composition—strongly predicts group performance. Empathy and anticipation further support the capacity to project future actions of teammates. Prior studies suggest that when team members have a strong understanding of each other’s skills and behavioral tendencies, they can make more accurate predictions about how teammates will react under different circumstances \citep{entin1999adaptive}. This ability to project is especially crucial in high-stakes environments, where split-second decisions rely on accurately anticipating teammates' responses. However, the literature also points to potential challenges; for example, personality traits such as overconfidence or reluctance to share information can hinder team SA and the accuracy of projections. Overconfident individuals may underestimate the need for feedback, leading to gaps in SA, while those less inclined to communicate openly may limit the ability of the team to form accurate mental models \citep{cooke2017communication}.

Coordination mechanisms in human-human teaming are also essential for optimizing collaboration and ensuring efficient task execution. A well-structured workflow outlines task organization, responsibility distribution, and interaction sequences. Information, decision, and incentive structures are additional coordination mechanisms that can enhance workflow performance. With enhanced information visibility, adaptive workflows can be designed to create flexibility in task execution. Decision structures that clarify team member responsibilities are increasingly recognized as vital to operating in dynamic environments \citep{entin1999adaptive}.

When teams are tasked with solving complex problems, coordination becomes more challenging due to the involvement of multiple agents with diverse skills, perspectives, and responsibilities. Such teams are often deliberately designed, and their success depends on navigating the increased complexity of communication and collaboration. 
\cite{rainey2015modeling} emphasize the importance of understanding team dynamics in multi-agent systems, suggesting that complexity arises not only from the individual characteristics of team members but also from their interactions and interdependencies. Teams performing complex tasks benefit from diversity in team member profiles (including skills, academic background, experiences, socio-economic status, and other factors).

\section{Outlook for Future Research in a New Era of AI Renaissance}
\label{sec:outlooks}

As AI agents continue to advance into increasingly proficient collaborators \citep{usaf2015autonomous}, human-AI teaming exemplifies a paradigm where humans and agents engage in iterative and multi-modal communications to pursue shared goals. Key phases of this collaboration include planning, facilitation, iteration, and testing. Maintaining optimal teaming dynamics in such settings demands not only coordination strategies but also conflict management to mitigate potential frictions and ensure effective task execution.

In this section, we mainly discuss future research by considering a simpler context: a single AI agent serving as a team member with a specific role. Building on Section \ref{sec:mapping}, we discuss four major human-AI teaming phenomena that are as yet underexplored and warrant further investigation: (1) unraveling the facts and mechanisms of the third level of human SA—projection, with a focus on human learning (Section \ref{sec:humanprojection}); (2) facilitating effective AI SA to enable AI (role) specification and AI learning (Section \ref{sec:aisa}); and (3-4) achieving human-AI shared SA by understanding and designing social (role) specification teams to facilitate team formulation and coordination (Section \ref{sec:teamformulation}), and by leveraging social learning for effective team maintenance, training, and long-term evolution (Section \ref{sec:teamcmaintenance}). 

\begin{figure}[t]
\centering
    \includegraphics[trim={37mm 72mm 88mm 30mm}, clip, width=\textwidth]{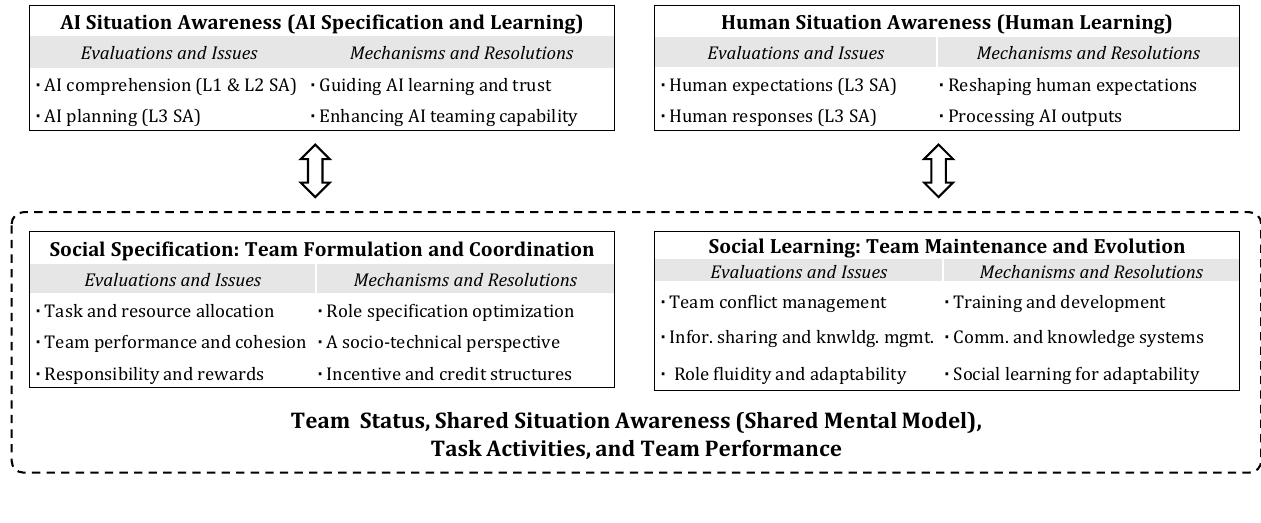}
\caption{Summary of Future Research Directions}
\label{fig:summary}
\end{figure}

We structure the four research streams by first outlining the desired outcomes for effective human-AI teaming as a guiding objective. We then explore the associated positive and negative aspects (i.e., tensions), addressing key issues, measurement challenges, and mechanisms and solutions needed to resolve potential obstacles in human-AI teaming. In general, we adopt a cautiously optimistic approach to motivating human-AI teaming, avoiding overly positive or negative expectations on AI capabilities  \citep{moser2024taking}. \Cref{fig:summary} provides a visual summary of the proposed research directions inspired by the extended team SA model, highlighting the significance of AI's contextual understanding and the team's adaptive capabilities.

\SingleSpacedXI
\DoubleSpacedXI

\subsection{Understanding Human Learning: Human Projection of AI Agents}
\label{sec:humanprojection}

Projection in team SA involves anticipating the future actions and consequences of both human and agent team members. As AI agents evolve, their increasing autonomy and adaptability necessitate a deeper understanding of how humans adjust their expectations and behaviors to these advanced collaborators.

\subsubsection{Expected Outcomes}

AI agents are expected to become active contributors/members rather than passive tools in team  \citep{vanneste2024artificial, dennis2023ai}. This shift envisions AI agents to process complex information and develop actionable insights to complement human creativity and decision-making. AI and human partnership can enhance problem-solving efficiency and reduce the cognitive workload on human team members. Particularly, \citet{salas2017situation} emphasize that trust among team members is essential to enhancing information sharing and reducing communication hesitancy. If trust in the AI team collaborator increases, it would lead to a more cohesive team dynamic. Trust in AI is expected to grow as its contributions become more accurate and contextually relevant, enabling smoother workflows and more confident task delegation. 
However, this integration brings challenges. Over-reliance on AI may suppress human creativity, learning, and critical thinking, as individuals increasingly defer to AI-generated solutions without adequate investigation \citep{fugener2021will, dell2023navigating}. Misaligned expectations, where AI outputs fail to meet team goals, can lead to frustration, diminished trust, and resistance to adoption. Furthermore, excessive dependence on AI for routine and complex tasks risks eroding human expertise, leaving teams vulnerable in scenarios where AI is unavailable or unreliable. Addressing these tensions is crucial for fostering sustainable and balanced collaboration in human-AI teams.

\subsubsection{Evaluations and Issues}

\noindent \underline{\textit{Altering Human Expectations:}} \ As discussed in Section \ref{sec:intelligentagentdef}, intelligent agents today are capable of interactions that are much more strongly anchored on the task context. Interactions over time at varying levels of intensity in task contexts will contribute to expectation formation for the human collaborator. Expectations represent a form of human learning that relates to human projection (Level 3) toward the external situation of the intelligent agent and the dynamics of teamwork \citep{frydman1982towards}. In iterative interactions, humans may change their expectations about the next input from their AI teammate, which is much more nuanced and complex than what has been studied in the literature on technology usage. The rich conversational interaction between AI and humans can lead to anticipative reasoning, planning, and discovery in both human and AI team members. Such a symbiotic interaction can result in better teamwork \citep{bayer2024interacting}. The new interaction styles can radically alter human evaluations of technology use where AI collaborations are not merely the cumulative or average result of individual stages of the interaction process as it is usually studied, but can often disproportionately be influenced by the most salient (``peak'') and final (``end'') moments of interactions (i.e., peak-end rule) \citep{fredrickson1993duration}. For example, an AI-generated output during a critical task phase or a resolution provided at the conclusion of an interaction can significantly enhance human expectations and satisfaction, even if earlier stages were less effective. Studying such complex interaction processes and outcomes poses an opportunity for IS researchers. Methodologically, constructs designed to capture the dynamics of interactions remain either insufficiently developed or underutilized in AI research. It is unclear if measurements used in human-human teaming contexts would fit well in examining human-AI teaming. Furthermore, a qualitative understanding of the learning process at the dyadic level presents new grounded theory development opportunities. Future research can take mixed-method approaches to describe the dynamic formation of expectations toward (1) AI outputs and reasoning, (2) cognitive loads, and (3) task and team performance. Learning within the human-AI dyad can also have an impact on human-human interactions within the team. Differing trust levels of AI among team members, task-AI fit across different team responsibilities, and the associated feedback loops present a complex challenging research domain that is an opportunity for new theory development in the information systems field. Accumulated theoretical knowledge in this domain can help organizations redesign tasks, calibrate task assignments, and reconfigure workflows.

\noindent \underline{\textit{Quantification of Human Responses:}} \ Following the canonical expectation-confirmation framework that is used to understand information systems continuance \citep{bhattacherjee2001understanding}, this kind of focal specific human responses relates to human projection (Level 3) toward external situation. The augmented AI capability may reshape humans' attitudes and learning behaviors \citep{lu2022promise}. It is imperative to further disambiguate human-AI interactions at the task level to gain new insights into how humans lean on the extended cognitive support for their work. Human responses can differ within a task as it pertains to idea generation, solution selection, conflict resolution, and task execution. In particular, incorporating the peak-end rule discussed earlier, it becomes critical to investigate how humans react to updates and augmentations provided by AI during task execution, particularly at moments of high significance or at the conclusion of interactions under each task type. These moments could have an outsized impact on humans' response to AI. Key metrics include response times, decision accuracy, and reliance on AI suggestions. Longitudinal studies can track changes in human (1) attitudes and (2) adaptation behavior toward AI over time, offering insights into the evolving trust and collaboration dynamics between humans and AI.

\subsubsection{Mechanisms and Resolutions}

\noindent \underline{\textit{Shaping Human Expectations:}} \ Human expectations for AI as a team member will likely be influenced through deployment and feedback mechanisms, which themselves be interdependent. Feedback mechanisms can be broadly categorized into (1) \textit{Descriptive feedback}: AI provides neutral, observational feedback providing information without judgment (e.g., summarizing data trends). Feedback supports situational awareness but may not directly guide decision-making. (2) \textit{Normative feedback}: AI makes goal-oriented suggestions emphasizing desired outcomes (e.g., recommending optimal actions). This feedback drives alignment with strategic objectives but may risk overdependence if the feedback is overly prescriptive. (3) \textit{Corrective feedback}: Error-oriented adjustments highlighting execution and reasoning inaccuracies or deviations (e.g., pointing out flaws in task assignments and execution). While corrective feedback enhances accuracy, it requires careful framing to avoid undermining human confidence. (4) \textit{Evaluative feedback}: Judgment-based assessments offering qualitative or quantitative evaluations (e.g., performance ratings). This feedback fosters accountability but can introduce biases if not transparently derived and delivered with a keen awareness of context. Humans absorb these feedback types through the integration of multiple channels (e.g., visual, auditory, sensory). Deployment also determines where the AI team member functions on the continuum of operant and operand resources. As an operand, AI functions as a passive entity requiring human activation. Whereas, as an operant, AI functions as an independent entity that makes independent judgments and takes actions on its own. The task interdependencies between AI agents and humans can become very fluid when human-AI interactions dynamically adjust along this continuum in reaction to situational awareness. We currently have no theoretical grounding to suggest optimal configurations for deployment and feedback mechanisms, let alone the interactions between the two or the potential possibility for dynamism in the mechanisms during task execution. For example, future studies can explore (1) constraints and ethical considerations related to feedback and deployment configurations, (2) impact on human learning and adaptability under different configurations, and (3) human expectations and experiences related to cognitive fluency and burden associated with AI agents moving from being an operand resource to an operant resource.

\noindent \underline{\textit{Processing Transformative AI Outputs:}} \ It is essential to understand how humans perceive, process, and learn from transformative AI outputs (e.g., suggestions, conclusions). An advanced AI teammate could break information bubbles and introduce novel perspectives for humans in their teamwork \citep{nguyen2014exploring}. This process involves analyzing how humans interpret AI-generated insights and the subsequent impact on their problem-solving approaches. 
There is a growing body of research that shows humans preferring AI for more objective decision contexts and preferring recommendations from another human for more subjective/complex decisions \citep{steyvers2024three}. In this vein, (1) to what extent such preferences are a function of current AI capabilities and human perceptions as opposed to longer-term behavioral preferences will remain an open question for the next few years. Transparency and explainability in recommendations can greatly influence the acceptance and utilization of AI recommendations in team processes. Furthermore, (2) to what extent humans believe in the neutrality of AI agents (i.e., not biased through managerial manipulations or organizational priorities that are not in alignment with human perception) can also be influential in how humans accept transformative inputs from AI agents. In addition to cognitive and perceptual outcomes, (3) studies that record interaction data to unpack the deliberation processes between humans and AI agents remain underexplored. This is particularly critical in the era of generative AI, as the performance of these agents heavily depends on how individuals utilize them and interpret the recommendations within the larger socio-technical context of their work. 

\subsection{AI Specification and Learning: AI SA of Humans and Team}
\label{sec:aisa}

Intelligent agents, such as those that recent advances in generative AI enable, possess advanced capabilities in perception, comprehension, and projection, which are key elements of SA. These abilities foster context-aware collaboration as AI is becoming more adept at interpreting human requirements, contextual nuances, and team activities. When combined with prescriptive capabilities, these intelligent agents can facilitate planning and reasoning to align with shared team goals, paving the way for more effective human-AI teaming \citep{teaming2022state}.

\subsubsection{Expected Outcomes}

The integration of AI agents with SA capabilities holds great promise for enhancing team collaboration. By interpreting instructions, understanding objectives, and anticipating needs, assigning specialized roles to AI and integrating it into appropriate workflows can enhance task efficiency, optimize resource allocation, and improve decision-making.  
The growing autonomy of AI agents, however, introduces challenges. When crafting AI with strong learning capabilities to understand tasks, process feedback, and adapt to dynamic contexts, it is crucial to recognize key limitations. AI specification problem in a team context spans several issues. Modern AI systems perform well in data-rich environments. While some contexts are naturally data-rich, in other contexts, deliberate design for data collection and aggregation is a must for developing reliable AI agents. In socially complex and subjective decision-making scenarios, utilization of AI agents requires striking a balance between the tacit expertise of humans and rational insights from the AI agents \citep{sharma2023towards}.

\subsubsection{Evaluations and Issues}

\noindent \underline{\textit{The AI Role Specification Challenge:}} \  
The understanding of team activities by AI can be broadly viewed as a challenge of role specification. The clarity and precision of the specifications governing both product and process characteristics in the team environment determine how well human-to-AI knowledge transfer is facilitated. Product characteristics encompass the attributes, requirements, and goals of the tasks or outputs the AI is expected to deliver, while process characteristics define workflows, rules, and interaction protocols that structure team collaboration with AI. Robust specifications are essential for ensuring that AI systems accurately perceive (Level 1) and comprehend (Level 2) their environment. When specification is imperfect, AI may struggle with accuracy, misinterpret ambiguous human inputs, or fail to adapt to changing objectives. Defining the parameters of AI role specification within the team is essential to enhancing its SA perception and the resulting interaction within the team. Future studies should investigate (1) how well AI systems identify task requirements and interpret human prompts, particularly in ambiguous scenarios involving both intentional and unintentional inputs. This includes examining the factors that influence AI’s accuracy in recognizing and adapting to dynamic team configurations. Additionally, (2) researchers should explore how AI systems process evolving objectives and integrate these changes into workflows to maintain alignment with team goals. These efforts will inform the development of more adaptive role specifications, ensuring that AI systems can respond effectively to the complexity and fluidity of human-AI collaboration \citep{teaming2022state}. 

\noindent \underline{\textit{AI Planning and Task Assignment Capabilities:}} \  
The capability of AI to autonomously plan and assign tasks reflects its Level 3 projection capability and highlights another critical dimension of the specification challenge, especially for complex and unstructured tasks. Effective task planning and assignment require a clear articulation of task characteristics and team capabilities, emphasizing the critical role of specification in defining AI's role specification. Research by \citet{fugener2022cognitive} demonstrates that AI delegating tasks to humans often yields superior outcomes compared to humans delegating tasks to AI. This finding underscores how well-defined specifications influence not only performance but also the dynamics of human-AI collaboration. Poorly defined specifications can result in suboptimal planning, resource misallocation, and difficulties in adapting to unforeseen circumstances \citep{ju2025collaborating}. Advanced algorithms that can assess task requirements and optimally distribute responsibilities within the team are essential to address these challenges. Future studies should explore (1) the alignment of AI-generated task plans with team objectives, focusing on how effectively these plans incorporate team strengths and address evolving task requirements. Additionally, researchers should (2) examine the efficiency and adaptability of task execution under AI leadership, particularly in dynamic, multi-stage projects. These efforts will inform the design of advanced planning systems that enhance AI's ability to allocate tasks effectively while maintaining smooth transitions and sustained collaboration within human-AI teams.

\subsubsection{Mechanisms and Resolutions}

\noindent \underline{\textit{Guiding AI Learning and Trust Decisions:}} \
From an AI learning perspective, determining what AI should learn and trust is critical to achieving effective human-AI teaming. The scale and scope of product and process contexts significantly influence AI’s learning speed, accuracy, and robustness. These contexts shape how AI systems interpret task requirements, team dynamics, and human feedback, making it essential to align AI learning processes with operational realities. A key challenge lies in identifying which human inputs AI should trust or disregard to optimize performance and collaboration, particularly in scenarios involving human-AI disagreements \citep{jorge2024should}. Emerging learning algorithms, such as inverse reinforcement learning and imitation learning, offer promising solutions by enabling AI to infer human goals and preferences from observed behaviors, enhancing adaptability in complex and ambiguous scenarios \citep{arora2021survey}. Future research should (1) develop frameworks to train AI systems in comprehending task objectives and team dynamics, focusing on how AI can integrate these insights into decision-making while responding effectively to evolving circumstances \citep{ghonasgi2024crucial}. Additionally, (2) investigate mechanisms that balance adaptability and objectivity in AI systems to prevent sycophantic behaviors, ensuring AI remains reliable and aligned with team goals even when faced with conflicting or ambiguous inputs \citep{sharma2023towards}. These efforts will help align AI learning processes with organizational values and strategic objectives \citep{shen2024towards,ji2023ai}, fostering synergy between human creativity and AI’s computational capabilities to enhance collective intelligence.

\noindent \underline{\textit{Enhancing AI Task Delegation and Team Collaboration:}} \
The ability of AI to autonomously plan and assign tasks represents a critical dimension of the AI learning challenge, especially for unstructured and complex tasks. The scale and complexity of product and process contexts significantly influence AI's ability to delegate tasks effectively while aligning with team objectives. Task delegation and collaboration require AI systems to not only understand task characteristics but also adapt to the capabilities and dynamics of team members. Advanced learning algorithms, such as contextual imitation learning and hierarchical reinforcement learning, offer promising approaches to enhance AI's ability to manage these challenges. By incorporating contextual awareness and operational constraints, these algorithms enable AI to generate task plans that are both efficient and collaborative. Future research should (1) focus on developing adaptive algorithms that dynamically adjust task plans based on interdependencies, operational constraints, and evolving team dynamics, ensuring alignment with team goals and enhancing task execution quality. Additionally, (2) investigate mechanisms for seamless task transitions between AI and human members, emphasizing protocols for continuity, clear communication, and role flexibility. These efforts will optimize AI's role in fostering team cohesion, improving performance outcomes, and enhancing the long-term adaptability of human-AI collaboration \citep{bansal2019updates, liu2024best}.

\subsection{Social Specification: Human–AI Team Formulation and Coordination}
\label{sec:teamformulation}

\begin{figure}[t]
\centering
    \includegraphics[trim={18mm 55mm 38mm 25mm}, clip, width=\textwidth]{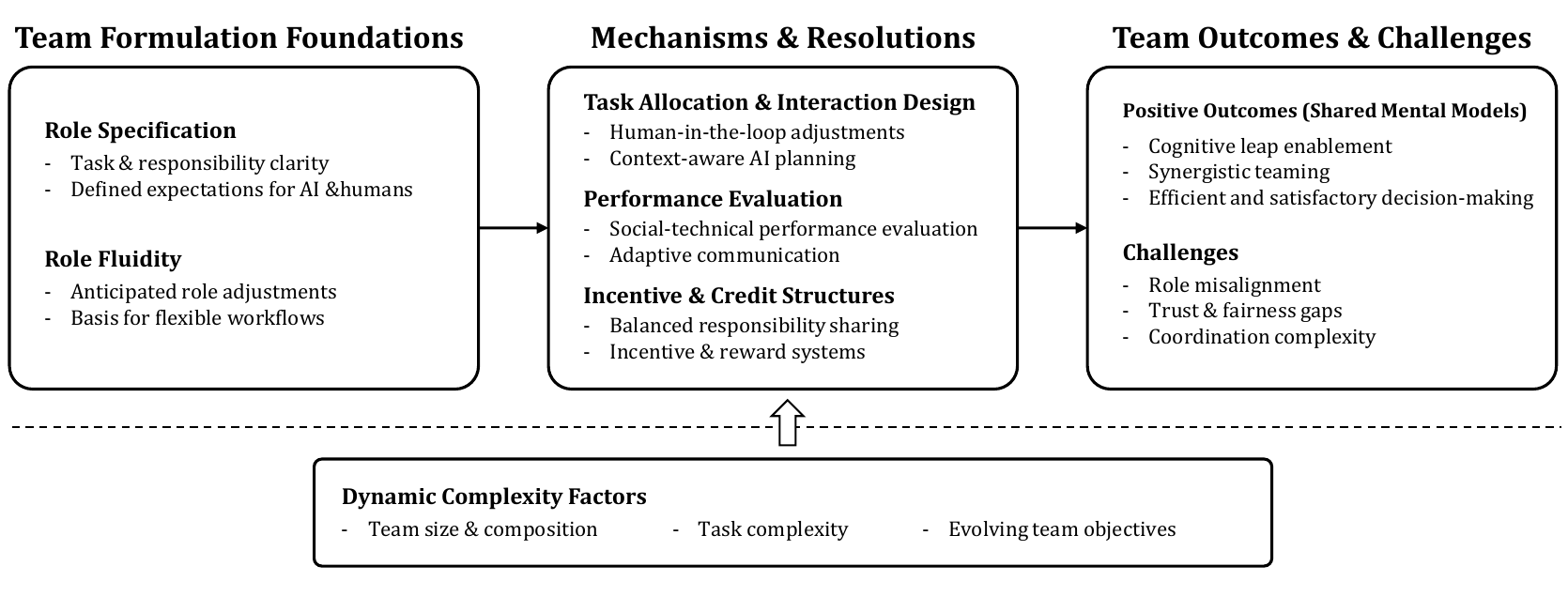}
\caption{A Framework for Team Formulation and Coordination in Human–AI Teams}
\label{fig:summarysec4.3}
\end{figure}

A human-AI team operates as a socio-technical system, requiring well-crafted social specifications to achieve cohesion and efficacy. Specifications such as defined roles, rules, and norms are needed to guide team formulation, coordination, and interaction. Particularly, as discussed before, human-AI teaming introduces new dynamics that demand modifications to engagement modes to integrate AI agents to achieve collective goals. Organizations that excel in managing these specifications will outperform their counterparts, leveraging both human and AI strengths effectively. Below, as summarized in Figure~\ref{fig:summarysec4.3}, we explore the expected outcomes, key evaluations and issues, and mechanisms for addressing opportunities and challenges in this domain, while proposing specific research directions for future exploration.

\SingleSpacedXI
\DoubleSpacedXI

\subsubsection{Expected Outcomes}

Human-AI teams can achieve synergy by combining AI's vast computational power with human intuition and expertise. This integration can enhance task execution, adaptability, and innovation. Optimized role specification allows team members, both human and AI, to leverage their strengths, resulting in faster decision-making, higher-quality outcomes, and greater satisfaction  \citep{panico2024value}.
Greater opportunities for research emerge in more complex and creative team tasks. While the focus has been on easing cognitive burden and augmenting cognitive capabilities, new research on social specifications that foster cognitive leaps is still scarce. We assert that cognitive leaps at current AI capabilities are only possible with humans in the loop \citep{choudhary2025human}. But AI support can greatly aid in humans achieving cognitive leaps in their tasks. 
By definition, cognitive leaps are discontinuous reasoning or insights as opposed to stepwise logical or linear reasoning. Helping humans connect concepts and data that at the surface seem unrelated can greatly aid cognitive leaps. Moreover, when the cognitive burden on routine reasoning tasks is eased, it enables humans to focus on more complex concepts and tasks. Social specification for human-AI teaming can hinder or encourage cognitive leaps through strategies that identify deeply hidden patterns in data, exploration of adjacent ideas and domains, synthesizing disjoint knowledge streams, and others. Achieving synergies is as much a technical problem (e.g., interaction design, data identification, model support) as it is a social problem (e.g., recognizing and rewarding human-AI innovations, autonomy for decision-making, ability to incorporate outside data sources).  This includes balancing accountability between human and AI members and developing mechanisms to value diverse contributions effectively \citep{teaming2022state, grote2024taming}.

\subsubsection{Evaluations and Issues}
\label{sec:socialspecializationevaluation}

\noindent \underline{\textit{Task Work, Role, and Resource Allocation Effectiveness:}} \  
Task and role allocation are critical for ensuring human-AI team synergy. Synergistic allocations effectively blend AI's computational strengths and human insights to achieve team objectives. The reason we identify social specification issues here is that for humans to effectively challenge/alter/recombine AI recommendations, they should be enabled to do so. That is, human team members who override or refine AI recommendations should be enabled and encouraged to take such actions. Yet, there should be a mechanism to evaluate their decisions against actual outcomes. The feedback mechanisms, however, need careful calibration lest they create risk-aversive behavior in the team. 
Future research should (1) explore methods for optimal team designs that encourage humans to contextualize AI recommendations, enable AI agents to discover deep patterns, and provide feedback mechanisms that improve both human and AI performance over time. In more creative and explorative task contexts, future research can (2) identify how AI can accelerate choice generation (such as prototypes and policy frameworks), and humans can weigh in to filter and refine the choice sets based on organizational context and competitive strategies. 

\noindent \underline{\textit{Team Performance and Cohesion:}} \ 
Trust, transparency, and shared understanding of specialized roles among team members influence team performance and cohesion in human-AI collaboration. Perceptions of AI’s capabilities and limitations play a significant role in shaping these dynamics. Advanced AI systems, while powerful, may introduce biases or errors that can undermine trust and fairness \citep{choi2024llm, lanz2024employees}. Moreover, multimodal AI features may inadvertently increase emotional harm risks if not carefully designed \citep{li2024perceived, gabriel2024ethics}. New team management approaches are needed to achieve team cohesion, trust levels, perceived fairness, and group identity. Addressing ethical concerns and minimizing biases in AI decision-making is critical for fostering trust and maintaining team morale \citep{kim2024ai}. Future research should thus (1) examine how trust and fairness perceptions evolve in human-AI teams and explore interventions to enhance cohesion. Researchers should also (2) investigate the role of emotional support mechanisms in reducing friction and improving collaboration \citep{strachan2024testing}, particularly in high-stakes or dynamic environments.

\noindent \underline{\textit{Responsibility and Rewards:}} \  Effective evaluation of responsibility assignments and reward systems is essential for understanding team motivation, fairness, and the successful integration of role specification in human-AI collaboration. The implications of human-AI teaming on reward and performance systems will be an important research focus for a number of reasons. First, reward systems have tended to recognize individual contributions, productivity, and effort. In most cases, such reward systems tend to work on the assumption that skills and effort are the primary drivers of performance. Human-AI teaming fundamentally alters this calculus where intuition, critical thinking, and creativity will have a large impact on performance (with AI providing the skills input in most cases). Second, the infusion of AI agents in teams will alter the reporting structure considerably. Predictions of the need for fewer managers in an AI-augmented world are almost uniformly acknowledged. At the minimum, this would lead to managers having a wider scope of oversight over multiple team members. The ability of managers to be able to recognize creative thinking among the human team members can be severely curtailed in such a context. Clearly, new performance assessment mechanisms will be needed to equitably assess team members. Clearly defined responsibilities that leverage the unique strengths of both human and AI members enhance accountability and efficiency by ensuring each member contributes according to their specifications. Conversely, poorly defined roles can lead to ambiguity, misaligned expectations, and inefficiencies, while imbalanced reward systems risk demotivating team members and undermining cohesion \citep{teaming2022state}. Third, any changes to performance measurement systems will instill feedback mechanisms that, in turn, dictate how human-AI teams perform. Therefore, assessing the efficacy of any performance measurement systems will be hard in the near term for any team. Future studies should (1) explore innovative methods to balance responsibility and rewards in human-AI teams, integrating role specification to ensure accountability and equitable recognition of contributions. (2) Mechanisms such as decentralized governance models and stakeholder negotiation frameworks \citep{grote2024taming} offer promising directions to enhance fairness and foster cohesion, ultimately optimizing team performance.

\subsubsection{Mechanisms and Resolutions}
\label{sec:socialspecializationmechanism}

\noindent \noindent \underline{\textit{Optimizing Role Specification and Adaptability:}} \ Role specification ensures that team members, both human and AI, can leverage their strengths to contribute effectively to team goals. In human-AI teaming, AI’s computational capabilities must complement human intuition and contextual understanding. However, rigid role structures may fail to adapt to dynamic task demands, necessitating more flexible approaches to role assignment \citep{marks2000performance}. Adaptive algorithms that adjust roles based on real-time performance and task requirements can enhance team effectiveness \citep{zhang2024rethinking, raisch2023combining}. We elaborate further on role fluidity in Section \ref{sec:teamcmaintenance}. Future research should (1) explore the development of frameworks for determining when AI should take on specific responsibilities, such as assisting in intermediate stages or making final decisions \citep{yin2025designing}. Additionally, researchers should (2) investigate how collaborative modes, such as parallel versus sequential workflows, impact team dynamics and outcomes \citep{chen2024large,choudhary2025human}.

\noindent \underline{\textit{Adopting A Socio-Technical Perspective on Performance Evaluation:}} \  
The socio-technical perspective provides a holistic framework for evaluating team performance by examining AI's impacts and key decisions across its lifecycle. It emphasizes assessing AI's suitability for tasks while addressing the needs of individuals, groups, and society \citep{schwartz2022towards}. By analyzing the interplay between human and technological factors, such as social dynamics, individual skills, tools, and organizational contexts, this perspective highlights drivers and barriers to team effectiveness. Addressing challenges like AI-induced job insecurity and maladaptive workplace behaviors is essential for fostering resilience and collaboration in human-AI teams \citep{yam2023rise}. Future studies should thus (1) investigate AI’s socio-psychological impacts on trust and motivation to improve team performance and cohesion. Moreover, fostering effective human-AI collaboration also requires addressing job security, emotional support, and adaptive teamwork. The perceived threat of AI replacing jobs undermines team cohesion and warrants urgent attention \citep{yam2023rise}. Game-theoretical models could be developed to provide insights into balancing human-AI roles for equitable outcomes \citep{panico2024value}. Skill-enhancing training and equitable work distribution are critical strategies for mitigating these risks. Similarly, experienced workers’ concerns about maintaining their unique contributions must be addressed, as advanced AI may reduce expertise gaps, potentially disadvantaging highly skilled individuals \citep{brynjolfsson2017can,simantov2024more}. Future studies should (2) develop AI systems with adaptive communication skills, such as interpreting social signals and recognizing personality traits \citep{li2024perceived, gabriel2024ethics}, to enhance trust and collaboration in human-AI teams.

\noindent \underline{\textit{Designing Effective Incentive and Credit Structures:}} \ Well-designed incentive and credit systems are vital for maintaining motivation and fairness in human-AI teams. Poorly designed structures can lead to disengagement, reduced morale, and diminished collaboration. Performance-based incentives, recognition programs, and credit systems that acknowledge contributions from both human and AI members are critical for fostering active participation and accountability \citep{teaming2022state}. Additionally, mechanisms to align individual goals with team objectives can promote collective responsibility and cohesion \citep{geng2022human}.
Future research should explore (1) innovative incentive models that balance financial and non-financial rewards, enhancing commitment and motivation in human-AI teams, as well as (2) the role of perceived equity in the distribution of responsibilities (perhaps more inline with creativity and insights needed in the tasks) and rewards, delving into how differences in team members' roles and contributions -- especially between humans and AI -- affect perceptions of fairness and team cohesion. 

\subsubsection{Discussion Extension: Considering Team and Task Complexities}

While our discussion has focused on simpler team structures, growing complexity in team size and composition presents new challenges for human–AI collaboration. Future research should examine how role and task allocation strategies scale in larger, more diverse teams. As complexity increases, maintaining clarity, adaptability, and alignment becomes more difficult. Investigating how task coordination performs across varying configurations can help develop robust frameworks that ensure effectiveness and coherence in increasingly dynamic environments.
Specifically, smaller teams with fewer members and one or two AI agents often benefit from streamlined communication and faster decision-making, making them effective for tasks requiring agility and close collaboration. Conversely, larger teams that bring a broader range of expertise face challenges such as coordination difficulties, slower decision-making, and increased risks of misalignment. This highlights the need for enhanced coordination mechanisms and adaptive workflows to maintain alignment and cohesion in larger teams. Moreover, teams involving multiple humans and multiple AI agents introduce dynamic interactions that heighten creativity but also require advanced strategies to manage diverse inputs and prevent cognitive overload. In addition to team size, task complexity significantly influences team formulation and coordination. Tasks with interdependencies across various objectives demand structured workflows that can dynamically adjust to AI-generated insights. For example, AI contributions in one task area may ripple into changes in related areas, requiring seamless transitions and alignment across team stages. Addressing these complexities underscores the critical role of role specification in managing interdependencies and ensuring alignment with overarching goals.

Considering these complexities, building on the discussions in \Cref{sec:socialspecializationevaluation,sec:socialspecializationmechanism}, future research should further (1) explore the optimal configurations of team size and composition for different task complexities, examining how these factors impact team dynamics and effectiveness \citep{wu2019large,yanai2024takes}, and (2) investigate mechanisms for managing task interdependencies and role transitions in dynamic, multi-functional human-AI teams. It also requires to strike a good balance between learning and productivity \citep{mayo2025coordination}. These efforts will provide insights into addressing the increasing complexity of modern human-AI teams while optimizing specification and coordination.

\subsection{Social Learning: Implications on Social Dynamics in Teams}
\label{sec:teamcmaintenance}

\begin{figure}[t]
\centering
    \includegraphics[trim={18mm 55mm 38mm 25mm}, clip, width=\textwidth]{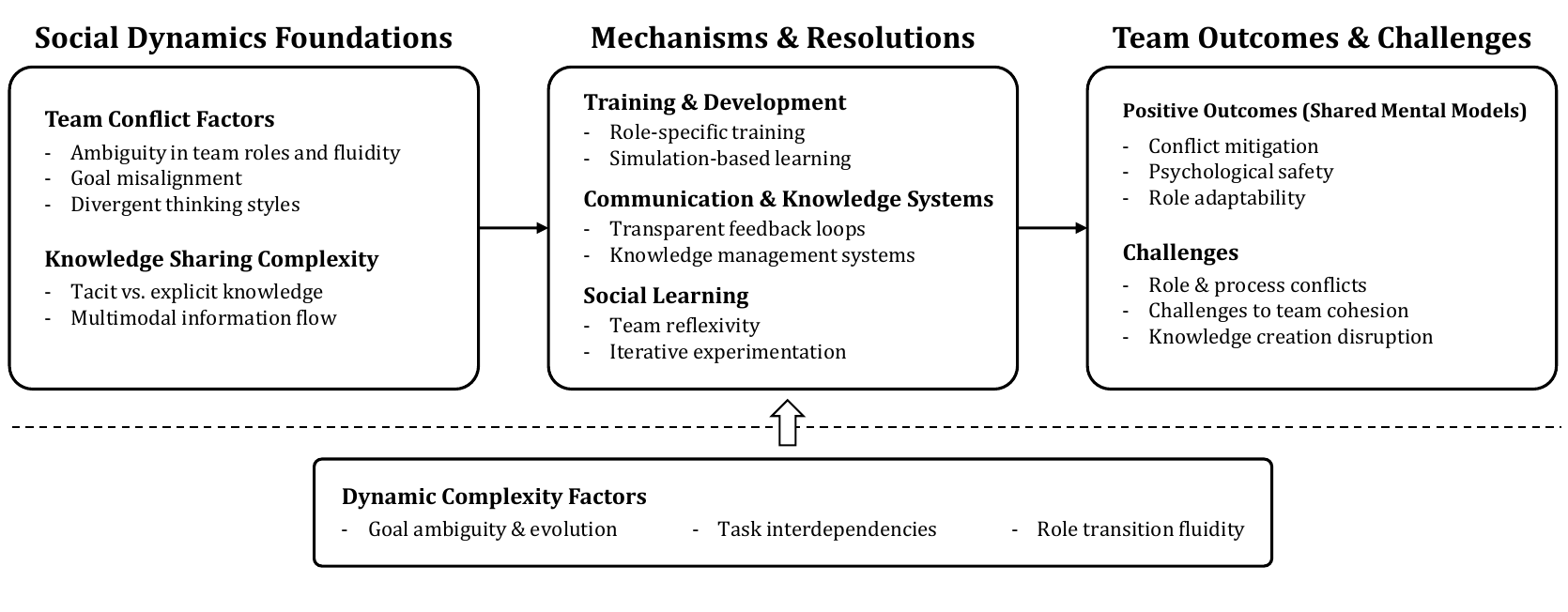}
\caption{A Framework for Team Maintenance and Evolution in Human–AI Teams}
\label{fig:summarysec4.4}
\end{figure}

Conflicts are inherent in all teams, whether human-only or human-AI. These conflicts typically arise from task, relational, and process-related dynamics \citep{de2013task}. In human-AI teams, such conflicts will continue to affect performance, but the generative or causal factors require new theoretical insights. Human-AI teams present new nuances, such as increased role fluidity—the ability to adapt and transition roles based on context and task requirements. Role fluidity can influence process clarity or conflict depending on how teams implement and manage their roles. If not managed early on, process conflict contributes to relational conflict. As AI agents take on more tasks, it is reasonable to expect changes to the shared mental models among team members. To the extent the mental models cohere, reductions in task conflicts are to be expected, which in turn may reduce process conflicts. It is, however, unclear as to whether the use of AI agents in teams interrupt the formation of shared mental models among humans or if they strengthen shared mental models. Below, we expand our discussion and outline key research directions, as summarized in Figure~\ref{fig:summarysec4.4}.

\subsubsection{Expected Outcomes}

Research on conflict in human-AI teams will need new theoretical lenses to study causative factors of conflict in such teams. The expected benefits of flat hierarchies and role fluidity have to be balanced against the potential harmful implications on team conflict. Extant research has raised the plausibility of increased interpersonal and status conflicts in flat organizational structures \citep{anicich2016bases}. Fluidity in roles often causes process conflict, which, if not managed, leads to relational conflicts. If the key benefits of human-AI teams are to be realized, teams need to more effectively manage such conflicts. Human-AI teams need to address imbalances in responsibility and rewards to sustain trust, engagement, and overall team effectiveness \citep{gabriel2024ethics}. We argue that human-AI teams need their leaders and team members to develop a stronger meta-understanding of process goals and roles as opposed to focusing on transactional level agreements on process and task execution. This, in our opinion, is the key differentiation in theorizing the social dynamics of human-AI teams.

\SingleSpacedXI
\DoubleSpacedXI

\subsubsection{Evaluations and Issues}
\label{sec:sociallearningevaluation}

\noindent \underline{\textit{Evaluating Conflict in Human-AI Teams:}} \  
Role fluidity enhances flexibility in task assignments and delegation but can lead to relational, process, and status-related conflicts if not well managed. Researchers should (1) explore how role fluidity leads to conflicts in human-AI teams and identify managerial actions that reduce such conflicts.  For example, AI agents can aid shared mental model formation through self-explanation and reflection, improving role clarity and minimizing conflicts. Attitudes toward both human and AI teammates, especially when performance varies due to task allocations, are critical. For example, \citet{simantov2024more} emphasize that employees’ trust in AI is linked to their performance rankings, making it vital to address dynamic perceptions of self-efficacy and collaboration willingness to maintain overall team resilience. Importantly, future research should also (2) evaluate how divergent and convergent thinking styles are balanced within human-AI teams to reduce task and process conflicts, focusing on the role of AI in fostering or disrupting this balance \citep{lanaj2018benefits}. This includes investigating how AI systems can facilitate social learning processes by enabling team members to integrate diverse perspectives, resolve conflicts constructively, and adapt roles dynamically for cohesive decision-making \citep{choi2024llm}.

\noindent \underline{\textit{Information Sharing and Knowledge Management in Human-AI Teams:}} \  Effective information sharing is fundamental to human-AI team coordination and performance, particularly for enabling role fluidity and adaptive learning. Knowledge reuse involves not only accessing archived information but also meaningfully applying it to new tasks, allowing the team to build on past experiences and insights. We draw upon Nonaka's dynamic theory of knowledge creation to identify key research directions for human-AI teams context \citep{nonaka1994dynamic}. Nonaka's model includes four knowledge interaction patterns: socialization (tacit–tacit), externalization (tacit–explicit), internalization (explicit–tacit), and combination (explicit–explicit). The inclusion of AI agents clearly has ramifications on the interaction patterns in human-AI teams as it relates to knowledge sharing and creation. Of the four patterns, perhaps the combination pattern is relatively well understood. AI agents today are highly capable of combining internal and external knowledge and synthesizing it for human consumption. Future research should (1) examine how human-AI teams can effectively organize to improve socialization, externalization, and internalization. For example, can AI agents impede tacit knowledge formation in teams since they may interrupt or replace socialization between team members? how can organizations enhance socialization in AI-enabled teams?  Additionally, (2) can the ability of AI agents to transform multimodal inputs (text, voice, video, etc.) accelerate knowledge creation through improved externalization? Similarly, (3) do interactions with AI agents during task completion impede or help internalization? Overall, human-AI teams will witness a disruption of the dyadic interactions in knowledge formation, and theoretical development in this domain is limited, making it difficult to reliably predict how organizational knowledge creation improves or deteriorates in the presence of AI agents.

\noindent \underline{\textit{Assessing Team Adaptability and Role Fluidity:}} \  
Adaptability and resilience are crucial for human-AI teams to sustain high performance in dynamic and challenging environments. Monitoring the stability and adaptability of team capabilities across multiple tasks allows for assessing how well teams handle disruptions, recover from setbacks, and maintain functionality under varying conditions. As mentioned before, role fluidity is central to these dynamics, as adaptive teams can reallocate responsibilities and adjust roles to meet changing demands. Continuous learning and feedback loops are essential for fostering such adaptability, enabling team members and AI systems to refine their performance and identify improvement opportunities. Future research should (1) investigate factors influencing ongoing team learning, such as the interplay of team dynamics, individual motivations, and organizational support systems \citep{woodruff2024knowledge}. Moreover, (2) exploring methods to create collaborative environments that enhance adaptability is critical. For example, simulation-based training and feedback loops can prepare teams for unforeseen challenges, while role rotation exposes team members to diverse responsibilities, fostering skill enhancement and flexibility \citep{de2022managing}. These efforts will reinforce social learning, promoting continuous improvement and ensuring long-term team effectiveness in dynamic and complex environments.

\subsubsection{Mechanisms and Resolutions}
\label{sec:sociallearningmechanism}

\noindent \underline{\textit{Training and Development for Collaboration:}} \  Developing targeted training strategies is essential for equipping human-AI teams to overcome cognitive limitations, enhance role fluidity, and foster effective collaboration. Training programs should focus on building shared mental models, active reasoning skills, and collaborative problem-solving techniques to enable the internalization of explicit knowledge and externalization of tacit knowledge. Promoting shared visions and values also helps establish a sense of belonging and commitment among team members \citep{beveridge2023theorizing}. Modules on AI literacy and role-specific training are crucial for helping human team members understand AI’s capabilities and limitations, facilitating effective teamwork \citep{kuhl2024investigating}. 
Incorporating adaptive learning techniques \citep{kaplan2021higher}, such as personalized learning paths, peer learning platforms, and interactive assessments, ensures that team members can dynamically adjust to the evolving functionalities and roles of AI within the team. Continuous professional development opportunities are vital to maintaining proficiency with advancements in AI technologies and team collaboration methods \citep{zhang2024omnificence}. Future research should (1) explore the investigation of different training and learning systems that improve externalization and internalization capabilities in teams. Additionally, (2) investigate the effectiveness of role-specific training programs in improving socialization (i.e., tacit-tacit knowledge creation), task allocation and enhancing role fluidity, particularly in complex, multi-functional human-AI teams, to optimize performance and adaptability.

\noindent \underline{\textit{Building Communication and Knowledge Management Systems:}} \  
Effective communication and robust knowledge management systems are essential for fostering role fluidity and sustaining social learning in human-AI teams \citep{straus1994does}. Transparent communication channels not only facilitate goal alignment and streamline feedback but also provide a foundation for structured discussions and experimentation, which are critical for fostering a culture of learning and adaptation. These channels must support open dialogue and enable teams to address conflicts constructively, ensuring that both human and AI members can align dynamically as tasks and conditions evolve. Knowledge management systems play a complementary role by capturing insights from team interactions and creating accessible, evolving repositories that enhance both immediate task performance and long-term adaptability \citep{grant1996toward}. Advanced archiving approaches, including automated tagging, context-aware indexing, and dynamic categorization, can ensure that knowledge is easily retrievable and reusable, allowing teams to internalize extant knowledge more quickly Future research should (1) investigate the impact of embedding real-time feedback mechanisms into communication frameworks, focusing on their ability to enhance team cohesion, alignment, and responsiveness during dynamic tasks. Additionally, (2) explore the potential of AI-driven knowledge management systems to actively support all four knowledge creation mechanisms identified in prior literature. These advancements will transform Agentic AI systems into dynamic partners for continuous learning and innovation, further strengthening the adaptability and resilience of human-AI teams.

\noindent \underline{\textit{Fostering Adaptability and Resilience through Social Learning:}} \  
Resilient human-AI teams thrive on structured social learning processes such as reflection, experimentation, and collaborative problem-solving, which enable them to co-adapt and address disruptions effectively. Team reflexivity -- collective reflection on objectives, strategies, and processes -- enhances problem identification and resolution, improving overall team adaptability \citep{schippers2014team}. Collaborative problem-solving fosters critical thinking and collective learning, equipping teams to navigate dynamic and uncertain environments \citep{van2006social}. These processes are integral to dynamically aligning roles between human and AI team members, strengthening team resilience. Particularly, simulated learning environments offer a practical and scalable strategy for accelerating these social learning processes. By providing safe, controlled spaces for experimentation, these environments enable teams to engage in iterative practice, refine workflows, and test strategies without the risks of real-world tasks \citep{salas2009using}. Simulation-based learning bridges theoretical knowledge and practical application, enhancing decision-making and adaptability in complex scenarios \citep{wei2024impact}. Iterative experimentation within such environments fosters mutual adaptation between humans and AI, supporting the development of shared understanding and improved collaborative efficiency. Psychological safety is a cornerstone of effective social learning. Creating environments where team members can share honest feedback and take calculated risks without fear of negative repercussions promotes openness, creativity, and innovation \citep{edmondson1999psychological}. Integrating simulated learning opportunities with feedback-rich environments ensures that human and AI agents can coevolve efficiently, addressing both operational efficiency and ethical alignment \citep{brinkmann2023machine}. Future research should (1) investigate how simulated learning environments can incorporate iterative experimentation -- such as real-time role rotation and collaborative problem-solving exercises -- to enhance role fluidity and team resilience. Additionally, (2) explore governance strategies that embed social learning processes, such as feedback loops and ethical audits, into human-AI workflows to improve adaptability, trust, and long-term alignment \citep{mokander2023operationalising}.

\subsubsection{Discussion Extension: Considering Goal and Task Dynamics}

As team goals evolve, maintaining and adapting human-AI teams becomes increasingly complex, requiring advanced strategies rooted in social learning, knowledge creation, and conflict management \citep{vashdi2013can}. While teams with clearly defined objectives benefit from focused workflows and aligned contributions, those with dynamic or ambiguous goals -- such as exploring innovative solutions or addressing shifting problem spaces -- demand enhanced role fluidity and iterative collaboration. Dynamic goals necessitate continuous adjustments to strategies and roles, as human creativity and intuition are required to contextualize AI-generated insights such as emerging patterns or alternative solutions. Effective social learning mechanisms are essential for aligning these contributions, enabling human and AI members to adapt cohesively to evolving objectives. Task complexity amplifies these challenges by introducing intricate interdependencies and unpredictable outputs from AI agents. The dynamic nature of AI contributions -- such as adaptive plans, evolving solutions, and segmented task objectives -- requires iterative refinement to ensure alignment with team goals. While social learning processes mentioned above are critical for maintaining alignment and continuity across workflows, managing these processes becomes increasingly challenging in the face of frequent role transitions and task ambiguities, highlighting the need for enhanced conflict management and adaptive training frameworks to support team resilience.

On top of the discussions in \Cref{sec:sociallearningevaluation,sec:sociallearningmechanism}, future research should (1) explore how advanced social learning frameworks can integrate conflict management and iterative feedback loops into team maintenance and training, ensuring that human-AI teams maintain alignment and cohesion as goals and tasks evolve. Additionally, (2) investigate mechanisms to enhance role fluidity by developing adaptive protocols for task transitions, interdependency management, and real-time adjustments, ensuring that both human and AI contributions remain dynamically aligned with team objectives. These efforts will provide actionable insights into sustaining the effectiveness, adaptability, and resilience of human-AI teams in increasingly complex and dynamic environments.

\section{Concluding Remarks}
\label{sec:conclude}

The disruptive innovation of AI agents compels us to perform a renewed examination of human–AI teaming. Through an extended team SA model, we present a comprehensive framework for examining key dimensions of human-AI teaming, including formulation, coordination, maintenance, and training. We also address issues arising from team complexity, such as variations in composition, size, goals, task dynamics, and long-term interactions. Ultimately, we emphasize the importance of revisiting human-AI teaming to foster a more dynamic and interactive relationship between humans and AI agents. Our work offers a forward-looking research agenda to guide future investigations in this evolving field.

Our discussion proposes the opportunities and challenges presented by the integration of AI systems into collaborative settings. On the one hand, AI offers immense potential for augmenting human capabilities, improving decision-making, and fostering innovation. On the other hand, unresolved tensions, such as trust and alignment considerations, underscore the need to design systems and processes that facilitate effective and sustainable collaboration and dynamic learning in teams. As such, future research should prioritize both technical development and the socio-behavioral foundations of teaming. 

We argue that successful human-AI teaming requires rethinking traditional team models, particularly in the context of shared goals and mutual contributions. As AI systems evolve from tools to intelligent agents, they bring unprecedented flexibility and autonomy while demanding new strategies, particularly around role specification and fluidity, to support effective team communication, coordination, and trust-building. The research agenda proposed in this paper offers a roadmap for addressing these challenges, with a focus on advancing the theoretical, practical, and ethical dimensions of human-AI collaboration. It is also highly relevant to the emerging \textit{multi-agent} literature, where, as demonstrated by recent studies (e.g., \citealp{cemri2025multi}), systemic frictions such as inter-agent misalignment and specification failures frequently lead to performance breakdowns across a wide range of contexts.

In conclusion, human-AI teaming is a critical frontier in team science, offering vast opportunities for innovation and productivity while raising questions about the nature of collaboration, decision-making, and shared accountability. By addressing the gaps and tensions identified in this paper, researchers and practitioners can shape a future where human-AI teams achieve their full potential, benefiting individuals, organizations, and society at large.

\SingleSpacedXII
\bibliographystyle{informs2014} 
\bibliography{0.main}

\end{document}